\documentclass[preprint,showpacs,preprintnumbers,amsmath,amssymb]{revtex4}

\usepackage{graphicx}
\usepackage{dcolumn}
\usepackage{bm}

\begin{document}


\title{A note on bounded entropies}

\author{Pierre-Olivier Amblard}
 \email{bidou.amblard@lis.inpg.fr}
  \homepage{http://www.lis.inpg.fr/pages_perso/bidou/BidouPerso.html}
\author{Christophe Vignat}%

\affiliation{%
Laboratoire des Images et des Signaux, Grenoble, CNRS, UMR 5083\\
BP46 38402 Saint-Martin d'H\`eres cedex , France
}%

\date{\today}

\begin{abstract}
The aim of the paper is to study the link between non additivity of some entropies and their boundedness. We propose an axiomatic construction of the entropy relying on the fact that entropy belongs to a group isomorphic to the usual additive group. This allows to show that the entropies that are additive with respect to the addition of the group for independent random variables are nonlinear transforms of the R\'enyi entropies, including the particular case of the Shannon entropy. As a particular example, we study as a group a bounded interval in which the addition is a generalization of the addition of velocities in special relativity. We show that Tsallis-Havrda-Charvat entropy is included in the family of entropies we define. Finally, a link is made between the approach developed in the paper and the theory of deformed logarithms. 
\end{abstract}

\pacs{02.50.-r, 05.20.-y}
\maketitle

\section{Motivations}

It is well known that the Shannon entropy of the couple of two independent random variables is the sum
of their respective Shannon entropies \cite{CoveT91}. This is also true for the R\'enyi entropy \cite{Reny66}. This fact comes from the well accepted axiom that the information given by two independent events should be the sum of their respective information. However,  this result is not true for all the entropies, and hence for all the definitions of the information content of an event. For example, the Tsallis-Havrda-Charvat (THC) entropy $H_\alpha$ \cite{HavrC67,Tsal88} with parameter $\alpha$ behaves as
\begin{eqnarray*}
H_\alpha (X,Y) = H_\alpha (X) + H_\alpha (Y) + (1-\alpha) H_\alpha (X)H_\alpha (Y)
\end{eqnarray*}
whenever $X$ and $Y$ are independent random variables. Therefore, the THC entropy is not additive.
Furthermore, it appears that THC entropy is bounded, from below if $\alpha<1$ and from above if $\alpha>1$.

Here, we try to generalize this observation by  constructing entropies that are bounded by arbitrary bounds. An elegant way of doing this is to suppose that the entropy lives in a group  $(\mathbb{X},\oplus)$, $\mathbb{X}$ being for example a bounded subset of the real line, $\oplus$ being the law of composition. 

In the following, starting from the postulates that entropy should be $\oplus$-additive for independent variables and that it is a generalized mean of the information of individual events, we obtain the general form of the entropy over a group isomorphic to $(\mathbb{R},+)$. We show as an example that the Tsallis-Havrda-Charvat entropy enters the framework presented here when the law $\oplus$ is a special case of the Lorentz law of velocities composition in special relativity. Furthermore, we establish the connection between the approach presented here and the approach developed by Kaniadakis and Naudts based on the notion of deformed logarithms.

\section{Entropy}

To obtain the general form of the entropy, we follow the lines of the derivation of the R\'enyi entropy in \cite{JizbA02}. The basic ingredients to build R\'enyi entropy are the following:
\begin{enumerate}
  \item the information of a couple of independent individual events is the sum of their respective information,
  \item the information of a random variable is a mean information of the individual information,
  \item the information is also additive for  independent random variables
\end{enumerate}
If the mean used is the so-called Kolmogorov-Nagumo mean, it can be shown that the entropies satisfying the three axioms are the R\'enyi entropies (including of course Shannon entropy as a particular case)\cite{JizbA02}. 
However, as mentionned in the introduction, some measures of information are not additive for a couple of independent random variables. We already gave the example of THC entropy. To generalize this idea, we are going to modify axiom 1, and replace it by two others. The first one stipulates that the entropy as a physical quantity must belong to a group. Hence we assume that entropy lies in a set $\mathbb{X}$ which has a group structure when endowed with an addition that we write $\oplus$. The second axiom will be a modified version of axiom 1 in which the usual addition is replaced by the $\oplus$ addition. Hence the set of axioms we are going to work with is
\begin{enumerate}
\item the information belongs to a group $(\mathbb{X},\oplus)$,
  \item the information of a couple of individual events is the $\oplus$-sum of their respective information,
  \item the information of a random variable is a mean information of the individual information,
  \item the information is also $\oplus$-additive for  independent random variables
\end{enumerate}
We now derive the general form of an entropy which satisfies this set of axioms.  

Let $X$ be a discrete random variable and $\{p_k\}_k$ its probability law, $p_k$ being the probability of an event labelled $k$. Let $ Y$ be another random variable, independent from $X$, whose probability law is denoted as  $\{q_k\}_k$. Let $I(p)$ be the information conveyed by a random variable of probability law $\{p_k\}_k$, and let $I^p_k$ be the information conveyed by an individual event labelled $k$.

According to the third axiom, the information $I(p)$ for a full law is a  mean of the individual information, 
and therefore there is a well behaved one-to-one and onto function $f : \mathbb{X} \rightarrow \mathbb{X}$  such that
\begin{eqnarray}
I(p)= f^{-1} \left( \sum_{k}^{} p_k f\left( I^p_k \right) \right).
\end{eqnarray}

We now seek admissible functions $f$ that ensure that the information is $\oplus$-additive for independent random variables. Taking the fourth axiom into account, or
\begin{eqnarray}
I(pq) = I(p)\oplus I(q),
\end{eqnarray}
requires that $f$ satisfies
\begin{eqnarray}
 f^{-1} \left( \sum_{kl}^{} p_k q_l f\left( I^p_k \oplus I^q_l \right) \right)  &=&
 f^{-1} \left( \sum_{k}^{} p_k f\left( I^p_k \right) \right)\oplus  f^{-1} \left( \sum_{l}^{} q_l f\left( I^q_l \right) \right).
 \label{constraintonf:eq}
\end{eqnarray}
To go further, we suppose now that we work on a group $(\mathbb{X},\oplus)$ isomorph to $(\mathbb{R},+)$. Let $M_\oplus : \mathbb{X} \longrightarrow \mathbb{R}$ be the associated morphism. We then have $a\oplus b = M_\oplus^{-1}(
M_\oplus(a)+M_\oplus(b) )$ for any $a$ and $b$ in $\mathbb{X}$. Using this in equation (\ref{constraintonf:eq}),
and introducing $g=M_\oplus\circ f^{-1}$, we get
\begin{eqnarray*}
g \left( \sum_{kl}^{} p_k q_l g^{-1}\left(  M_\oplus(I^p_k) + M_\oplus(I^q_l) \right)  \right)  &=&
 g\left( \sum_{k}^{} p_k g^{-1}\left( M_\oplus( I^p_k) \right) \right) + g \left( \sum_{l}^{} q_l g^{-1}\left( M_\oplus( I^q_l)  \right) \right) .
\end{eqnarray*}
This expression is true whatever $p$ and $q$, and thus true for a uniform law $q$, that is $I^q_l=I, \forall l$. If we set $J=M_\oplus(I)$ and  $ J^p_k =M_\oplus(I^p_k) $ ,  we get
\begin{eqnarray*}
g\left(\sum_{k}^{} p_k g^{-1}\left( J^p_k + J \right)  \right)  &=& 
 g\left( \sum_{k}^{} p_k g^{-1}\left( J^p_k \right) \right) + J,
\end{eqnarray*}
an equation that has been solved already for $g$ (see for example  \cite{JizbA02} and references therein), the only solutions of which can be written as 
\begin{eqnarray}
g^{-1}(z+J)=a(J) g^{-1}(z) +g^{-1}(J) .
\label{solgene:eq}
\end{eqnarray}
Since $g^{-1}(z+J)=g^{-1}(J+z)$, we can write $a(J) g^{-1}(z) +g^{-1}(J)=a(z) g^{-1}(J) +g^{-1}(z) $
or $(a(J)-1)/g^{-1}(J)=(a(z)-1)/g^{-1}(z)=\gamma$. Inserting this into equation (\ref{solgene:eq}) leads to
the following functional equations
\begin{eqnarray*}
g^{-1}(J+z) &=& g^{-1}(J) + g^{-1}(z) \mbox{ for } \gamma=0 \\
a(J+z) &=& a(J)a(z)  \mbox{ for } \gamma\not=0.
   \end{eqnarray*}
The first case leads to $g^{-1}(x)=cx$ whereas the second one imposes to function $a$ to be an exponential.  Then,  we can write in this case $g^{-1}(x) = (2^{(1-\alpha)x}-1)/\gamma$.
Therefore, since $g^{-1}=f\circ M_\oplus^{-1}$, for $\gamma=0$ we get $f(x)=cM_\oplus(x)$ and for $\gamma\not=0$ we obtain $f(x)=(2^{(1-\alpha)M_\oplus(x)}-1)/\gamma$. 
Using the second axiom, the information of individual events is $\oplus$-additive for independent events, or
\begin{eqnarray}
I^{pq}_{kl}=I^p_k \oplus I^q_l
\end{eqnarray}
for individual independent events of probability $p_k$ and $q_l$.   Hence, for individuals events, $I$ is a morphism between $(\mathbb{R}^{+*},\times)$ and $(\mathbb{X},\oplus)$: it sends the usual multiplication onto the $\oplus$ addition. We write $M_\times$ this morphism to  obtain 
\begin{eqnarray*}
I(p) &=& M_\oplus^{-1} \left(  \sum_k p_k M_\oplus( M_\times (1/p_k))  \right) \mbox{ for } \gamma=0, \\
I(p) &=& M_\oplus^{-1} \left(\frac{1}{1-\alpha} \log_2\left[  \sum_k p_k 2^{(1-\alpha)M_\oplus( M_\times (1/p_k))}   \right] \right) \mbox{ for } \gamma\not=0.
\end{eqnarray*}
Now, we have $M_\times : (\mathbb{R}^{*+},\times) \longrightarrow (\mathbb{X},\oplus)$ and $M_\oplus : (\mathbb{X},\oplus) \longrightarrow (\mathbb{R},+)$. Therefore $M_\oplus\circ M_\times : (\mathbb{R}^+,\times) \longrightarrow (\mathbb{R},+)$ is the morphism sending multiplication onto the usual addition. In other words, $M_\oplus\circ M_\times (p)\propto \log (p)$.
Therefore, in the last equations, we choose $M_\oplus\circ M_\times (p) = \log_2 (p)$ and we end up with
\begin{eqnarray*}
I(p) &=& M_\oplus^{-1} \left(  -\sum_k p_k \log_2(p_k) \right) \mbox{ for } \gamma=0, \\
I(p) &=& M_\oplus^{-1} \left(\frac{1}{1-\alpha} \log_2\left[  \sum_k p_k^\alpha  \right] \right) \mbox{ for } \gamma\not=0.
\end{eqnarray*}
If we impose the usual additivity, that is to say $M_\oplus \propto Id$, we recover the Shannon or R\'enyi entropies. 
Note that this result is quite poor in itself since  the entropies that are $\oplus$-additive are nonlinear transformations of the usual R\'enyi or Shannon entropies. As a consequence, the maximizing probability laws are the same since the nonlinear transform is a morphism and as such is one-to-one and onto.

\section{Lorentzian addition and its application for bounded entropies}

In order to get bounded entropies, we can work into a bounded subset of the real line. A possibility relies on a generalization of the velocity composition law in special relativity. 
Let us consider the Lorentz addition 
\begin{eqnarray*}
I_1\oplus I_2 = \frac{I_1 + I_2 - I_1 I_2 \left( \frac{1}{h_+}+\frac{1}{h_-} \right)}{1-\frac{I_1 I_2}{h_- h_+}}
\end{eqnarray*}
This law, in its symmetrical form $h_-=-h_+$, was used by L. Nottale in his theory of scale relativity 
\cite{Nott93} to define a generalized composition of $\log$ scales. The asymmetrical form was developed by Dubrule\&Graner \cite{DubrG96}, and used by Dubrulle in his theory of finite size scaling \cite{Dubr00}. We further used it to define finite size scale invariant stochastic processes \cite{BorgAF05}.

The group considered here is $([h_-,h_+],\oplus)$ where $h_-<0$ and $h_+>0$. We recognize the velocity composition law in special relativity when $-h_-=h_+$ is taken to be the speed of light. 
 Furthermore, a semi-infinite case is obtained if one of the two bounds is infinite, or, the law becomes 
\begin{eqnarray*}
I_1\oplus I_2 &=& I_1 + I_2 -  \frac{ I_1 I_2}{h_-} \mbox{ if  }  h_+ \longrightarrow + \infty\\
I_1\oplus I_2 &=& I_1 + I_2 -  \frac{ I_1 I_2}{h_+} \mbox{ if  }  h_- \longrightarrow - \infty
\end{eqnarray*}
We then recover   the usual addition by letting the remaining finite bound to reach infinity.
Furthermore, the identity element of the group is 0 since $I \oplus 0=0 \oplus I=I$.
To get the associated morphism,  we  assume differentiability of $M_\oplus$, write $
M_\oplus(I\oplus J) = M_\oplus(I)+M_\oplus(J)$
and  
differentiate this expression  with respect to $J$ and then set $J=0$ to obtain the following differential equation
\begin{eqnarray*}
M_\oplus'(x) = \frac{M_\oplus'(0)}{(1-x/h_-)(1-x/h_+)},
\end{eqnarray*} 
whose solution reads    
\begin{eqnarray*}
M_\oplus(x)& =&M_\oplus'(0)\frac{h_-h_+}{h_+-h_-}\log \left(\frac{1-x/h_+}{1-x/h_-}\right)  .
\end{eqnarray*}
If the two bounds go to infinity, we easily check that the morphism is proportional to the identity, hence recovering the usual addition. 
By properly setting the free parameter $M_\oplus'(0)$, the inverses can be written as follows 
\begin{eqnarray*}
M_\oplus^{-1}(x) &= & h_+ \left( 1-2^{-x/h_+}  \right) \mbox{ if } h_- \longrightarrow - \infty     \\
&= &h_- \left( 1-2^{-x/h_-}  \right)  \mbox{ if }  h_+ \longrightarrow  + \infty \\
&=& h_-h_+\frac{1-2^{\frac{(h_+-h_-)x}{h_-h_+}} }{h_--h_+2^{\frac{(h_+-h_-)x}{h_-h_+}} }  \mbox{ else }
\end{eqnarray*}
Note that for the semi-infinite cases  it is implicitly assumed that 
\begin{itemize}
  \item for the case $h_- \longrightarrow -\infty $,  $h_+>0$ so that $\mathbb{X}=]-\infty, h_+[$
  \item for the case $h_+ \longrightarrow +\infty $,  $h_-<0$ so that $\mathbb{X}=] h_-, +\infty[$
\end{itemize}

It is clear that in the unbounded case $h_\pm \longrightarrow \pm \infty$ we recover either the Shannon entropy or the R\'enyi entropy, since when the bounds go to infinity, the morphism converges to  the identity.
In the case of the semi-infinite addition however, we get
\begin{eqnarray*}
I(p)= h_\pm \left( 1- \left[  \sum_k p_k^\alpha \right]^{-\frac{1}{(1-\alpha )h_\pm}} \right)
\end{eqnarray*}
where we recover as a special case the Tsallis-Havrda-Charvat entropy for $h_\pm (1-\alpha)=-1$, since then the entropy reads
\begin{eqnarray*}
I(p)= \frac{1}{(1-\alpha )} \left(   \sum_k p_k^\alpha -1 \right)
\end{eqnarray*}
The case $\alpha<1$ corresponds to an entropy bounded from below by $h_-=1/(\alpha-1)<0$ whereas $\alpha>1$ leads to an entropy bounded above by $h_+=1/(\alpha-1)>0$. 

This derivation of the THC entropy clearly shows the interpretation  of parameter $\alpha$ as a bound on the THC entropy. But as seen also from the derivation, $\alpha$
 is really  an additional free parameter in the analysis. It is not a physical parameter. Rather, the physics enters the entropy {\it via} the bounds $h_\pm$. $\alpha$ is thus a parameter used to perform the right analysis of the physical system, as it is used in the multifractal analysis.

In the bounded case, the entropy reads
\begin{eqnarray*}
I(p)=h_-h_+  \frac{ 1- \left(\sum_k p_k^\alpha \right)^{\frac{h_+-h_-}{(1-\alpha )h_+ h_-}}}
{h_- -h_+ \left( \sum_k p_k^\alpha \right)^{\frac{h_+-h_-}{(1-\alpha )h_+ h_-}}}
\end{eqnarray*}
and the three parameters can be let as they are. However, it should be pointed out again that the bounds are physical quantities, part of a thermodynamics that could be derived from this entropy. 
But in the same way we adopted to recover THC entropy in the semi-infinite case, we can eliminate one parameter by relating the bounds to $\alpha$. For example, in order to get a simpler expression of the entropy, we can choose
\begin{eqnarray*}
\frac{h_+-h_-}{h_+ h_-} = 1-\alpha
\end{eqnarray*}
We then have the freedom to keep one of the bounds as a free parameter. For example, if we keep $h_+$ as a free parameter, we obtain for the entropy the following expression
\begin{eqnarray*}
I(p)=h_+  \frac{ 1- \sum_k p_k^\alpha }
{1 - (1+(1-\alpha) h_+ )  \sum_k p_k^\alpha }
\end{eqnarray*}
an entropy bounded below by $h_- = h_+/ (1+(1-\alpha) h_+ ) $, and above by $h_+$.

\section{the link with the framework of deformed logarithms and exponentials}

\begin{figure}[h]
\includegraphics{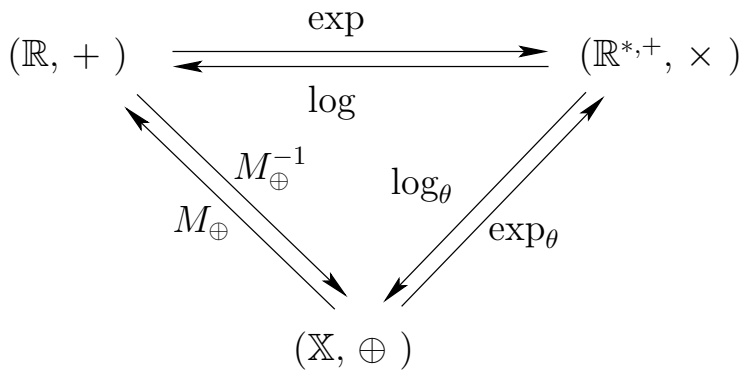}
\caption{Equivalent groups used and the morphisms associated to link them.}
\label{diagramme:fig}
\end{figure}

In a series of papers \cite{Kani01,Kani02:pre,KaniLS04,KaniLS04:arxiv}, Kaniadakis and co-workers, as well as Naudts \cite{Naud02,Naud04}, have developed the framework of entropy based on deformed logarithms. The idea in this framework is to consider that the entropy is an average quantity, and thus should write
\begin{eqnarray*}
S(p) = \sum_k p_k \Lambda_\theta ( p_k )
\end{eqnarray*}
where $\Lambda_\theta ( p_k )$ is the information associated to the single event labelled $k$, of probability $p_k$. Parameter $ \theta$ rules the behavior of $\Lambda$. Function $\Lambda_{\theta}$ is called a $\theta$-deformed
logarithm if it satisfies certain properties \cite{Naud02}. An interesting interpretation has been given by Kaniadakis in \cite{Kani02:pre} in terms of group theory. His point of view is similar to the idea we present here, and is based on the  diagram in figure (\ref{diagramme:fig}). This diagram explicits the equivalence between the three groups of interest, namely $(\mathbb{R},+),(\mathbb{R}^{*,+}, \times)$ and $(\mathbb{X},\otimes)$.  The functions appearing on the arrows represent the morphisms linking the groups. The notations $\log_\theta$ and $\exp_\theta$ represent the $\theta$-deformed logarithm and exponential, respectively. 

The difference between the approach presented here and that of Kaniadakis and Naudts, despite the conceptual approach,  relies especially in the fact that space $\mathbb{X}$ can be different from $\mathbb{R}$, and that we use the more general Kolmogorov-Nagumo generalized mean to define the entropy.

\section{Discussion}

The result in itself
\begin{eqnarray*}
I(p) &=& M_\oplus^{-1} \left(  -\sum_k p_k \log_2(p_k) \right) \mbox{ for } \gamma=0 \\
I(p) &=& M_\oplus^{-1} \left(\frac{1}{1-\alpha} \log_2\left[  \sum_k p_k^\alpha  \right] \right) \mbox{ for } \gamma\not=0
\end{eqnarray*}
is not very impressive. This is due to the fact that the group considered is isomorphic to $(\mathbb{R},+)$. It implies also that the maximising probability laws of the bounded entropies presented here are the same as the maximizing probability laws of the initial entropy. More interesting is the fact that THC entropy is a special case of entropies that are non additive (in the usual sense) but satisfy
\begin{eqnarray*}
I(pq) = I(p)+I(q) -\frac{1}{h_+} I(p)I(q) 
\end{eqnarray*}
for independent variables. This observation could open new twists in non-extensive statistical mechanics.
However, the relevance of the framework presented in this note remains to be proved on practical examples.
Let us finally mention that the work presented here includes the results of Dukkipati {\it et. al.} \cite{Dukk05}
concerning the non-generalizability of Tsallis entropy  using generalized mean.


\begin{thebibliography}{99}
\bibitem{HavrC67} J. Havrda, F. Charvat, Kybernetica, 3, 30, 1967.
\bibitem{Tsal88} C. Tsallis, Possible extension of the Boltzmann-Gibbs statistics, J. Stat.  Phys., 52, (1988), 479--487.


\bibitem{Nott93} L. Nottale, Fractal space-time and microphysics, World Scientific, Singapore, 1993.



\bibitem{Dubr00} B. Dubrulle, Finite scale size invariance, Eur. Phys. J, B", 14, 757--771, 2000.

\bibitem{BorgAF05} P. Borgnat, P.~O.  Amblard and P. Flandrin, Scale invariances and Lamperti transformations for stochastic processes, Journal of Physics, A : math. gen. , 38, 2081--2101, 2005.



\bibitem{JizbA02} P. Jizba, T. Arimitsu, The world according to R\'enyi: thermodynamics of multifractal systems, Ann. Phys., 312 (2004), 17-59.
\bibitem{CoveT91} T.M. Cover, J. A. Thomas, Elements of information theory, Wiley, 1993.
\bibitem{Reny66} A. R\'enyi, Calculus of probability, Dunod,1966. 
\bibitem{DubrG96} B. Dubrulle, F. Graner, Possible statistics of scale invariant systems, J. Phys. II, France, 6 (1996), 797-816.
\bibitem{Kani02:pre}  G. Kaniadakis, Statistical mechanics in the context of special relativity, Phys. Rev. E, 056125 (2002).
\bibitem{Kani01} G. Kaniadakis, Nonlinear kinetics underlying generalized statistics, Physica A, 296, pp405--425, (2001).
\bibitem{CzacN02} M. Czachor, J. Naudts, Thermostatistics based on Kolmogorov-Nagumo averages: unifying framework for extensive and nonextensive generalizations, Phys. Lett. A, 298, pp369--374, (2002)
\bibitem{Naud02} J. Naudts, Deformed exponentials and logarithms in generalized thermostatistics, Physica A, 316, pp323--334, (2002).
\bibitem{Naud04} J. Naudts, Generalized thermostatistics based on deformed exponential and logarithmic functions, Physica A, 340, pp32--40, (2004)
\bibitem{KaniLS04} G. Kaniadakis, M. Lissia, A.~M.  Scarfone, Deformed logarithms and entropies, Physica A, 340, pp41--49, (2004).
\bibitem{KaniLS04:arxiv} G. Kaniadakis, M. Lissia, A.~M.  Scarfone, Two-parameter deformations of  logarithm, exponential and entropy: a consistent framework for generalized statistical mechanics, arXiv:cond-mat/0409683, (2004).
\bibitem{Dukk05} A. Dukkipati, N.N. Murty and S. Bhatnagar, Nongeneralizability of Tsallis entropy by mean of Kolmogorov-Nagumo averages under pseudo-additivity,  arXiv:math-ph/0505078, (2005).
\end{thebibliography}
 \end{document}